\documentclass[12pt]{article}
\usepackage{epsfig}
\usepackage{latexsym}
\usepackage{bm}
\def\b{\beta}

\def\th{\theta}
\def\k{\kappa}
\def\RR{{\mathcal R}}
\def\II{{\mathcal I}}

\newcommand{\lam}{\lambda}
\newcommand{\del}{\partial}

\newcommand{\absv}[1]{\left|#1\right|}

\newcommand{\gtsim}{\mathrel{\hbox{\raise0.2ex
\hbox{$>$}\kern-0.75em\raise-0.9ex\hbox{$\sim$}}}}
\newcommand{\ltsim}{\mathrel{\hbox{\raise0.2ex
\hbox{$<$}\kern-0.75em\raise-0.9ex\hbox{$\sim$}}}}
\newcommand{\lw}[1]{\smash{\lower2.0ex\hbox{#1}}}

\newcommand{\Lag}{{\cal L}}

\newcommand{\kslash}{k\kern-0.5em\raise 0.14ex\hbox{/}}
\makeatletter

\@addtoreset{equation}{section}
\makeatother
\begin{document}
\begin{flushright}
June~16, 2005
\end{flushright}
\vskip10mm
\begin{center}
{\Large\bf Sphalerons in the NMSSM}\\[24pt]
{\bf Koichi~Funakubo$^{a,}$\footnote{e-mail: funakubo@cc.saga-u.ac.jp},
Akira~Kakuto$^{b,}$\footnote{e-mail: kakuto@fuk.kindai.ac.jp}
Shuichiro Tao$^{a,}$\footnote{e-mail: tao@higgs.phys.kyushu-u.ac.jp}
and Fumihiko~Toyoda$^{b,}$\footnote{e-mail: ftoyoda@fuk.kindai.ac.jp}}
\vskip20pt
{\it $^{a)}$Department of Physics, Saga University, Saga 840-8502 Japan}
\vskip 0.2 cm
{\it $^{b)}$School of Humanity-Oriented Science and Engineering,
Kinki University, Iizuka 820-8555 Japan}
\end{center}
\vskip6mm
\centerline{\bf Abstract}
\vskip4mm
\begin{center}
\begin{minipage}[t]{0.9\textwidth}
\baselineskip=14pt
We study sphaleron solutions in the next-to-minimal supersymmetric standard model.
We find that the boundary condition on the singlet field at the origin of the radial coordinate
is of Neumann type, while the other boundary conditions are of Dirichlet type.
The sphaleron energy takes almost the same value as in the MSSM for wide range of parameters,
in spite of the negative contribution from the cubic term in the Higgs potential.
\end{minipage}
\end{center}
\baselineskip=16pt
\section{Introduction}
Since the discovery of the sphaleron solution\cite{Manton}, 
unsuppressed transition of the baryon number at high temperatures has been recognized 
to play an important role in early universe.
It yielded new possibilities of matter generation, baryogenesis through leptogenesis and 
electroweak baryogenesis\cite{EWBreview}.
The latter requires that the sphaleron process decouples just after the electroweak phase transition,
in order to protect the generated baryon asymmetry from washout by sphaleron process in equilibrium.
This sphaleron decoupling condition is expressed as
$\Gamma_{\rm sph}(T)\simeq T e^{-E_{\rm sph}/T}< H(T)$,
where $E_{\rm sph}$ is the static energy of the sphaleron solution and $H(T)$ is the Hubble parameter
at temperature $T$. In the standard model with one Higgs doublet, this condition can be cast into
the form of $v_C/T_C>1$, where $v_C$ is the expectation value of the Higgs field at the transition
temperature $T_C$, where the gauge symmetry is spontaneously broken.
Thus, the knowledge of energy of the sphaleron solution in a model viable for 
electroweak baryogenesis is indispensable to qualify the model by examining the sphaleron decoupling
condition.
The sphaleron solution was originally found in the 4-dimensional $SU(2)$ gauge theory with
one Higgs doublet\cite{KlinkhammerManton}.
The minimal standard model, however, cannot afford to generate the baryon asymmetry at the
electroweak phase transition with an acceptable Higgs mass. It was also pointed out that
the CP violation in the KM matrix is insufficient to provide the chiral charge as the source of
baryon number. Hence, one must extend the standard model for successful electroweak baryogenesis.
In particular, models with more bosonic fields are expected to admit the strongly first-order
phase transition. Among those models are the MSSM with a light stop and the two-Higgs-doublet model.
The sphaleron solutions have been found in the two-Higgs-doublet model\cite{2HDM} and in
the MSSM with finite-temperature corrections\cite{mssm}.
The model with one Higgs doublet and one singlet is among those extensions of the standard model
for which a sphaleron solution has been found\cite{doublet-singlet}.\par
We recently found that the Next-to-MSSM (NMSSM), which contains a singlet superfield in addition to the MSSM,
can have strongly first-order phase transition, with heavy stops and Higgs bosons whose masses and couplings 
are consistent with results of collider experiments so far\cite{nmssm-pt}.
Although the NMSSM has more parameters than the MSSM, there are more constraints on these parameters
which are absent in the MSSM. For example, the electroweak vacuum is not always the absolute minimum
of the Higgs potential. This is because the cubic terms containing the singlet scalar can easily generate
minimum of the potential other than the electroweak vacuum. We observed that the vacuum condition, which ensures that
the electroweak vacuum be the absolute minimum of the effective potential, together with the spectrum condition
on the Higgs scalars, restricts allowed parameters in the model\cite{FunakuboTao}.
In order to determine the parameter sets which are suited for the baryogenesis, one must know
the sphaleron solution and its energy in this model.
In the MSSM-limit, where the vacuum expectation value of the singlet goes to infinity, the sphaleron solutions 
are expected to have the same profile as the MSSM. When the expectation value is of order of the weak scale,
we expect new features of the phase transitions, but even the existence of the sphaleron solution is not
obvious. Our aim is to find the sphaleron solution and to study its energy in the case of 
weak-scale expectation value of the singlet field.\par
Since the sphaleron solution is a saddle-point configuration, it is difficult to find it by solving
the full equations of motion. Instead, we usually adopt the ansatz including the parameter along
the noncontractible loop in the configuration space\cite{KlinkhammerManton}.
The ansatz is constructed in such a way that the configuration corresponds to the vacuum at
the loop parameter $\mu=0$ and $\pi$, while the configuration at $\mu=\pi/2$ is expected to
be the highest-energy one corresponding to the sphaleron.
In order to ensure that the configuration is the sphaleron, one must check that the configuration
does satisfy the full equation of motion and that the fluctuation spectrum around the configuration
does contain only one negative mode. In practice, these procedure are complicated and has been 
done only in the one-doublet model\cite{Akiba}.
Here we extend the ansatz including the noncontractible loop to the NMSSM with $U(1)$-gauge
sector being turned off, and write down the static spherically symmetric equations of motion 
for the configuration at $\mu=\pi/2$.
Then we solve the equations of motion and evaluate the energy along the noncontractible loop
around the solution to ensure that it is the highest-energy configuration along the loop.\par
This paper is organized as follows.
In \S\ref{sec:ansatz}, we extend the ansatz of Klinkhammer and Manton for the configurations
along the noncontractible loop to the NMSSM. Then we derive the equations of motion for  the
sphaleron configurations. 
Several numerical solutions for the equations are shown in \S\ref{sec:num-sol} for parameter sets
corresponding to the light-Higgs and the heavy-Higgs scenarios.
Section~\ref{sec:discussions} is dedicated to concluding remarks.
The derivation of the asymptotic solutions, together with the boundary conditions, are
summarized in the Appendix.
\section{Ansatz and equations of motion}\label{sec:ansatz}
\subsection{noncontractible loop}
As discussed in \cite{Manton}, any static configuration of finite-energy doublet scalar $\Phi(\bm{x})$
defines a map from $S^2$ spanned by the spatial coordinates $(\theta,\phi)$ to $S^3=SU(2)$ which
characterizes the field at $r=\infty$ as a unitary transformation of the vacuum $\pmatrix{0\cr 1}$.
A one-parameter family of such configurations connecting vacua, which cannot be contracted to
a point, is constructed by realizing the family of maps as a map from $S^3$ to $S^3$.
Among such one-parameter families, that of the highest symmetry is expected to form the least 
energy set of configurations. The configuration with the highest energy along the parameter will
be a saddle-point configuration with one negative mode.\par
The Manton's ansatz for the noncontractible loop in the one-doublet model is
\begin{eqnarray}
 \Phi(\mu,r, \th, \phi) &=&
 {v\over{\sqrt2}}\left\{\left(1-h(r)\right)\pmatrix{ 0\cr e^{-i\mu}\cos\mu} + h(r)U(\mu,\th,\phi)\pmatrix{0\cr 1}\right\}, 
                                    \label{eq:noncont-phi}\\
 A_i(\mu,r,\th,\phi) &=& -{i\over g}f(r)\del_i U(\mu,\th,\phi)\,U^{-1}(\mu,\th,\phi),   \label{eq:noncont-A}
\end{eqnarray}
where
\begin{equation}
 U(\mu,\th,\phi) = \pmatrix{
  e^{i\mu}(\cos\mu-i\sin\mu\cos\th) & e^{i\phi}\sin\mu\sin\th \cr
  -e^{-i\phi}\sin\mu\sin\th  & e^{-i\mu}(\cos\mu+i\sin\mu\cos\th) }.      \label{eq:noncont-U}
\end{equation}
Here $\mu\in[0,\pi]$ and the configurations are reduced to the vacuum at $\mu=0$ and $\pi$.
As for the NMSSM, we adopt the following ansatz for the doublet Higgs fields, $\Phi_d$ and $\Phi_u$,
and the singlet $n$,
\begin{eqnarray}
 \Phi_d(\mu,r,\th,\phi) &=& 
 {{v_d}\over{\sqrt2}}
   \left\{\left(1-h_1(r)\right)\pmatrix{e^{i\mu}\cos\mu\cr 0} + h_1(r) U(\mu,\th,\phi)\pmatrix{1\cr 0}\right\}  \nonumber\\
 &=&
 {{v_d}\over{\sqrt2}}\pmatrix{ e^{i\mu}\left(\cos\mu -ih_1(r)\sin\mu\cos\th\right) \cr -e^{-i\phi}h_1(r)\sin\mu\sin\th },
                                               \label{eq:noncont-Phid} \\
 \Phi_u(\mu,r,\th,\phi) &=&
 {{v_u e^{i\rho}}\over{\sqrt2}}
  \left\{ \left(1-h_2(r)\right)\pmatrix{0\cr e^{-i\mu}\cos\mu} + h_2(r) U(\mu,\th,\phi)\pmatrix{0\cr1} \right\} \nonumber\\
 &=&
 {{v_u e^{i\rho}}\over{\sqrt2}}
   \pmatrix{ e^{i\phi}h_2(r)\sin\mu\sin\th \cr e^{-i\mu}\left(\cos\mu +ih_2(r)\sin\mu\cos\th\right)},\label{eq:noncont-Phiu}\\
 n(\mu, r,\th,\phi) &=& {{v_n e^{i\varphi}}\over{\sqrt2}} k(r).       \label{eq:noncont-n}
\end{eqnarray}
Now the profile functions to be solved are $f(r)$, $h_1(r)$, $h_2(r)$ and $k(r)$.\par
The lagrangian of the gauge-Higgs sector of the NMSSM is given by
\begin{equation}
 \Lag = 
 -{1\over4}F_{\mu\nu}^a F^{a\mu\nu} + \left(D_\mu\Phi_d\right)^\dagger D^\mu\Phi_d
 + \left(D_\mu\Phi_u\right)^\dagger D^\mu\Phi_u + \del_\mu n^*\del^\mu n 
 - V_0(\Phi_d, \Phi_u, n),               \label{eq:nmssm-lagrangian}
\end{equation}
where the Higgs potential is given by
\begin{eqnarray}
 V_0(\Phi_d, \Phi_u, n) \!\!\!&=&\!\!\!
 m_1^2\Phi_d^\dagger\Phi_d + m_2^2\Phi_u^\dagger\Phi_u +m_N^2\absv{n}^2
 -\left(\lambda A_\lambda n\Phi_d\Phi_u+{\k\over3}A_\k n^3 +\mbox{h.c.}\right) \nonumber\\
 &&
 +{{g_2^2+g_1^2}\over8}\left(\Phi_d^\dagger\Phi_d - \Phi_u^\dagger\Phi_u\right)^2
 +{{g_2^2}\over2}\absv{\Phi_d^\dagger\Phi_u}^2    \nonumber\\
 &&
 + \absv{\lambda}^2\absv{n}^2\left(\Phi_d^\dagger\Phi_d + \Phi_u^\dagger\Phi_u\right)
 + \absv{\lambda\Phi_d\Phi_u + \k n^2}^2,        \label{eq:nmssm-V0}
\end{eqnarray}
The terms in the first line of (\ref{eq:nmssm-V0}) are the soft supersymmetry-breaking terms,
those in the second line comes from the $D$-term potential, which are the same as those in the MSSM,
and those in the last line are the $F$-terms, which are peculiar to the NMSSM.
Here we adopt the same notation as in \cite{FunakuboTao}.\par
All the parameters in the potential with mass dimensions are to be determined once one specifies the
breaking mechanism of the supersymmetry and the mass scale relevant to it.
Instead of specifying supersymmetry-breaking mechanism, we express them in terms of
the vacuum expectation values (VEVs) of the Higgs fields by requiring that the tree-level potential
has vanishing first derivatives at the prescribed vacuum parametrized by the VEV of the doublet $v_0$,
the ratio of the doublet VEVs $\tan\b$ and the VEV of the singlet $v_n$.
As shown in \cite{FunakuboTao}\footnote{Since we are working with $g_1=0$, we should set 
$m_Z=m_W$ in the results in \cite{FunakuboTao}.},
the derivatives of $V_0$ with respect to the CP-even order parameters vanish at the vacuum, if
the soft masses are given by
\begin{eqnarray}
 m_1^2 &=& 
  \left(R_\lambda-{1\over2}\RR v_n\right)v_n\tan\b - {1\over2}m_W^2\cos(2\b)
    -{{\absv{\lambda}^2}\over2}\left(v_0^2\sin^2\b+v_n^2\right),       \\
 m_2^2 &=& 
  \left(R_\lambda-{1\over2}\RR v_n\right)v_n\cot\b + {1\over2}m_W^2\cos(2\b)
    -{{\absv{\lambda}^2}\over2}\left(v_0^2\cos^2\b +v_n^2\right),       \\
 m_N^2 &=&
  \left(R_\lambda-\RR v_n\right){{v_0^2}\over{v_n}}\sin\b\cos\b + R_\k v_n - 
  {{\absv{\lambda}^2}\over2}v^2 - \absv{\k}^2 v_n^2,
\end{eqnarray}
while those with respect to the CP-odd order parameters vanish, if
\begin{equation}
 I_\lambda = {1\over2}\II v_n.   
\end{equation}
Here we defined the following parameters which are independent of phase convention,
\begin{eqnarray}
 R_\lambda &=& {1\over{\sqrt2}}{\rm Re}\left(\lambda A_\lambda e^{i(\rho+\varphi)}\right),\qquad
 I_\lambda = {1\over{\sqrt2}}{\rm Im}\left(\lambda A_\lambda e^{i(\rho+\varphi)}\right), \nonumber\\
 R_\k &=& {1\over{\sqrt2}}{\rm Re}\left(\k A_\k e^{3i\varphi}\right),   \nonumber\\
 \RR &=& {\rm Re}\left(\lambda\k^* e^{i(\rho-2\varphi)}\right),   \qquad
 \II = {\rm Im}\left(\lambda\k^* e^{i(\rho-2\varphi)}\right).
\end{eqnarray}
We use these relations to express the static energy functional which is 
\begin{equation}
 E[\Phi_d, \Phi_u,n] =
 \int d^3\bm{x}\left[ {1\over4}F_{ij}^a F_{ij}^a + (D_i\Phi_d)^\dagger D_i\Phi_d 
   + (D_i\Phi_u)^\dagger D_i\Phi_u + \absv{\del_i n}^2
   + V_0(\Phi_d, \Phi_u, n) \right],           \label{eq:nmssm-energy-functional}
\end{equation}
in terms of the profile functions.
Upon inserting the ansatz for the noncontractible loop, (\ref{eq:noncont-phi}), (\ref{eq:noncont-A}),
(\ref{eq:noncont-Phid}), (\ref{eq:noncont-Phiu}) and (\ref{eq:noncont-n}), into 
(\ref{eq:nmssm-energy-functional}), we obtain
\begin{eqnarray}
 \lefteqn{ E[f,h_1,h_2,k](\mu)}    \nonumber\\
 &=&
 {{4\pi v_0}\over{g_2}}\int_0^\infty d\xi \Biggl\{
  4\sin^2\mu \left[ (f')^2+{{2f^2(1-f)^2}\over{\xi^2}}\sin^2\mu\right]  \nonumber\\
 &&\qquad
 +{{\cos^2\b\sin^2\mu}\over2}\Bigl[ \xi^2 (h_1')^2+ 2\bigl( h_1^2(1-f)^2+(1-h_1)^2 f^2\cos^2\mu  \nonumber\\
 &&\qquad\qquad\qquad\qquad\qquad\qquad   
                  -2h_1(1-h_1)f(1-f)\cos^2\mu \bigr) \Bigr]       \nonumber\\
 &&\qquad
 +{{\sin^2\b\sin^2\mu}\over2}\Bigl[ \xi^2 (h_2')^2+ 2\bigl( h_2^2(1-f)^2+(1-h_2)^2 f^2\cos^2\mu  \nonumber\\
 &&\qquad\qquad\qquad\qquad\qquad\qquad   
                  -2h_2(1-h_2)f(1-f)\cos^2\mu \bigr) \Bigr] 
     +{{N^2\xi^2}\over2}\left(k'\right)^2    \nonumber\\
 &&\qquad
 + {A\over8}\xi^2\sin^2\b\cos^2\b \left[ \left( h_1^2+h_2^2-2k(h_1h_2-1)-2\right)\sin^2\mu + (k-1)^2\right]  \nonumber\\
 &&\qquad
 +{{C_\RR N^2\xi^2}\over4}\sin\b\cos\b\left[ 2k(k-1)(h_1h_2-1)\sin^2\mu + (k-1)^2 \right]  \nonumber\\
 &&\qquad
 +{{\xi^2}\over{32}}[(h_1^2-1)\cos^2\b-(h_2^2-1)\sin^2\b]^2 \sin^4\mu   \nonumber\\
 &&\qquad
 +{{\xi^2}\over{12}} \sin^2\b\cos^2\b\sin^2\mu\cos^2\mu (h_1-h_2)^2 \nonumber\\
 &&\qquad
 +{{C_\lam \xi^2}\over4}\sin^2\b\cos^2\b \sin^2\mu
      \left[ (h_1h_2-1)^2\sin^2\mu - {1\over3}(h_1-h_2)^2(2+\sin^2\mu) \right]        \nonumber\\
 &&\qquad
 +{{C_\lam N^2\xi^2}\over4} (k^2-1)\left[ (h_1^2-1)\cos^2\b+(h_2^2-1)\sin^2\b\right]\sin^2\mu   \nonumber\\
 &&\qquad
  + {{C_\k N^4\xi^2}\over4}(k^2-1)^2 - {{B N^2\xi^2}\over6}(2k^3-3k^2+1)  \Biggr\}     \nonumber\\
 \!\!\!&=&\!\!\!
 {{4\pi v_0}\over{g_2}}\left( c_2\sin^4\mu + c_1\sin^2\mu + c_0 \right),    \label{eq:E-mu}
\end{eqnarray}
where $\xi=g_2v_0r$ and the primes on the profile functions denote the derivatives with respect to $\xi$.
We introduced dimensionless parameters defined by
\begin{eqnarray}
 A &=&  {{(R_\lam-\RR v_n/2)v_n}\over{m_W^2\sin\b\cos\b}} = {{\hat m^2}\over{m_W^2}}
 = {4\over{g_2^2v_0^2}}\left(m_{H^\pm}^2-m_W^2+{{\lambda^2}\over2}v_0^2\right),  \nonumber\\
 N &=& {{v_n}\over{v_0}},     \qquad
 B = {{R_\k N}\over{g_2^2 v_0}},           \label{eq:def-NMSSM-params}\\
 C_\RR &=& {\RR\over{g_2^2}},  \qquad
 C_\lam = {{\absv{\lambda}^2}\over{g_2^2}},  \qquad C_\k = {{\absv{\k}^2}\over{g_2^2}},      \nonumber
\end{eqnarray}
and the coefficients in the last expressions are defined by
\begin{eqnarray}
 c_2 \!\!\!&=&\!\!\!
 \int_0^\infty d\xi \Biggl[
  {{8f^2(1-f)^2}\over{\xi^2}} -\cos^2\b f(1-h_1)(f-2h_1+fh_1) -\sin^2\b f(1-h_2)(f-2h_2+fh_2)  \nonumber\\
  &&\qquad
  +{{\xi^2}\over{32}}\left( (h_1^2-1)\cos^2\b-(h_2^2-1)\sin^2\b \right)^2
  -{{\xi^2}\over{12}}\sin^2\b\cos^2\b (h_1-h_2)^2    \nonumber\\
  &&\qquad
   + {{C_\lam\xi^2}\over4}\sin^2\b\cos^2\b\left( (h_1h_2-1)^2-{1\over3}(h_1-h_2)^2\right) \Biggr],  \label{eq:coeff-sin-mu-4}\\
 c_1 \!\!\!&=&\!\!\!
 \int_0^\infty d\xi \Biggl[
  4(f')^2 + {{\cos^2\b}\over2}\left( \xi^2(h_1')^2+2(h_1-f)^2\right) 
            + {{\sin^2\b}\over2}\left( \xi^2(h_2')^2+2(h_2-f)^2\right)    \nonumber\\
 &&\qquad
  + {A\over8}\xi^2\sin^2\b\cos^2\b\left( h_1^2+h_2^2-2k(h_1h_2-1)-2\right)    \nonumber\\
 &&\qquad
  + {{C_\RR N^2\xi^2}\over2}\sin\b\cos\b\,k(k-1)(h_1h_2-1)
  + {{1-2C_\lam}\over{12}}\xi^2\sin^2\b\cos^2\b (h_1 - h_2)^2      \nonumber\\
 &&\qquad
  + {{C_\lam N^2\xi^2}\over4}(k^2-1)\left( (h_1^2-1)\cos^2\b+(h_2^2-1)\sin^2\b \right) \Biggr],    \label{eq:coeff-sin-mu-2}\\
 c_0 \!\!\!&=&\!\!\!
 \int_0^\infty d\xi\, \xi^2 \Biggl[
  {{N^2}\over2}(k')^2 + {A\over8}\sin^2\b\cos^2(k-1)^2 + {{C_\RR N^2}\over4}\sin\b\cos\b (k-1)^2  \nonumber\\
 &&\qquad
  + {{C_\k N^4}\over4}(k^2-1)^2 - {{BN^2}\over6}(2k^3-3k^2+1) \Biggr].     \label{eq:coeff-sin-mu-0}
\end{eqnarray}
In the presence of the CP violation, the $\II$-dependent term in the potential is proportional to
$\sin\mu\cos\mu\cos\theta$, which vanishes upon integration over $\theta$ to yield the static energy of the
configuration. This might suggest that an ansatz without spherical symmetry reduces the potential energy
compared to the spherically symmetric one in the CP-violating case, but such an ansatz with less symmetry
will increase the gradient energy. Hence we expect that the spherically symmetric ansatz yields
the least energy configuration for each $\mu$, as long as the CP violation is not so large.\par
The equations of motion for the sphaleron are derived by applying the variational method to
the energy functional (\ref{eq:E-mu}) with $\mu=\pi/2$.
After solving the equations, we check that the configuration is at least locally maximum at
$\mu=\pi/2$ along the noncontractible loop by calculating the coefficients $c_2$ and $c_1$.
That is, the coefficients calculated with the solution must satisfy $2c_1+c_2>0$, for which
$E''(\mu=\pi/2)<0$.
\subsection{equations of motion}
The static energy functional at $\mu=\pi/2$ is given  by
\begin{eqnarray}
 \lefteqn{ E[f,h_1,h_2,k] }    \nonumber\\
 \!\!\!&=&\!\!\!
 {{4\pi v_0}\over{g_2}}\int_0^\infty d\xi\Biggl\{
  4\left[(f')^2+{{2(f-f^2)^2}\over{\xi^2}}\right] + {{N^2}\over2}\xi^2(k')^2    \nonumber\\
 &&\qquad
 +{1\over2}\cos^2\b\left[\xi^2(h'_1)^2+2(1-f)^2h_1^2\right] 
 +{1\over2}\sin^2\b\left[\xi^2(h'_2)^2+2(1-f)^2h_2^2\right]       \nonumber\\
 &&\qquad
 +{A\over8}\sin^2\b\cos^2\b\,\xi^2(h_1^2+h_2^2+k^2-2h_1h_2k-1)   \nonumber\\
 &&\qquad
 +{{C_\RR N^2}\over4}\sin\b\cos\b\,\xi^2\left( 2k(k-1)h_1h_2-(k^2-1)\right)    \nonumber\\
 &&\qquad
 +{{\xi^2}\over{32}}\left[(h_1^2-1)\cos^2\b-(h_2^2-1)\sin^2\b\right]^2  \nonumber\\
 &&\qquad
 +{{C_\lambda}\over4}\xi^2 \left[\sin^2\b\cos^2\b (h_1^2-1)(h_2^2-1)
 +  N^2 (k^2-1)(h_1^2\cos^2\b+h_2^2\sin^2\b-1) \right] \nonumber\\
 &&\qquad
 +{{C_\k N^4}\over4}\xi^2(k^2-1)^2 
 - {{B N^2}\over6} \xi^2 (k-1)(2k^2-k-1)   \Biggr\}.    \label{eq:E-sph-xi}
\end{eqnarray}
Applying the variational method to this functional, we obtain the equations of motion 
for the profile functions:
\begin{eqnarray}
 f''(\xi) \!\!\!&=&\!\!\!
 {2\over{\xi^2}}f(1-f)(1-2f) - {{1-f}\over4}\left(h_1^2\cos^2\b+h_2^2\sin^2\b\right),  \label{eq:EOM-f-xi}\\
 \lefteqn{ h''_1(\xi)+{2\over\xi}h'_1(\xi) }    \nonumber\\
 \!\!\!&=&\!\!\!
 {{2(1-f)^2h_1}\over{\xi^2}} +
 {A\over4}\sin^2\b(h_1-h_2k) + {{C_\RR N^2}\over2}\tan\b\,h_2k(k-1)   \label{eq:EOM-h1-xi}\\
 &&\!\!\!
 + {{h_1}\over8}\left((h_1^2-1)\cos^2\b-(h_2^2-1)\sin^2\b\right)
 + {{C_\lam}\over2}\left( (h_2^2-1)\sin^2\b + N^2(k^2-1) \right)h_1,    \nonumber\\
 \lefteqn{ h''_2(\xi)+{2\over\xi}h'_2(\xi) }  \nonumber\\
 \!\!\!&=&\!\!\!
 {{2(1-f)^2h_2}\over{\xi^2}} +
 {A\over4}\cos^2\b(h_2-h_1k) + {{C_\RR N^2}\over2}\cot\b\,h_1k(k-1)   \label{eq:EOM-h2-xi}\\
 &&\!\!\!
  - {{h_2}\over8}\left((h_1^2-1)\cos^2\b-(h_2^2-1)\sin^2\b\right)
 +{{C_\lam}\over2}\left( (h_1^2-1)\cos^2\b + N^2(k^2-1) \right)h_2    \nonumber\\
 \lefteqn{ k''(\xi)+{2\over\xi}k'(\xi) }   \nonumber\\
 \!\!\!&=&\!\!\!
 {A\over{4N^2}}\sin^2\b\cos^2\b(k-h_1h_2) + {{C_\RR}\over2}\sin\b\cos\b\left(h_1h_2(2k-1)-k\right) \nonumber\\
 &&
 + {{C_\lam}\over2}(h_1^2\cos^2\b+h_2^2\sin^2\b-1)k + C_\k N^2(k^2-1)k - Bk(k-1).    \label{eq:EOM-k-xi}
\end{eqnarray}
In order to solve these equations, we must specify the boundary conditions for the profile functions.
We relegate the derivation of asymptotic behaviors of the solution to the appendix
and present only the result here.
As shown in the appendix, the profile functions with finite energy functional satisfy the boundary conditions
\begin{eqnarray}
 && f(0) = h_1(0) = h_2(0) =0, \qquad k'(0) = 0,       \label{eq:bc-0}\\
 && f(\infty)= h_1(\infty) = h_2(\infty) = k(\infty) = 1.  \label{eq:bc-infty}
\end{eqnarray}
Note that only $k(\xi)$ at $\xi=0$ must satisfy the boundary condition of Neumann type, while
the others satisfy those of Dirichlet type.
\section{Numerical solutions}\label{sec:num-sol}
In order to implement numerical study of the sphaleron solutions by solving the equations of motion
(\ref{eq:EOM-f-xi}) -- (\ref{eq:EOM-k-xi}) with the boundary conditions (\ref{eq:bc-0}) and
(\ref{eq:bc-infty}), we adopt the relaxation method which is suited for a boundary problem of
ordinary differential equations.
The NMSSM has more parameters than the MSSM, but some of the parameters are constrained
by the vacuum condition and the Higgs spectrum condition\cite{FunakuboTao}.
The model with the weak scale expectation value of the singlet scalar exhibits new features
with respect to the Higgs spectrum and the phase transitions\cite{nmssm-pt}, which are absent in
the MSSM. In particular, there are four types of phase transitions at finite temperatures.
Although the sphaleron solution which is relevant to the decoupling of the sphaleron process
at the first-order electroweak phase transition is a configuration which mediates the symmetric phase 
and the broken phase at the transition temperature, we here study sphaleron configurations
with the tree-level potential for parameters satisfying the tree-level vacuum condition.\par
In the case of the weak scale expectation value of the singlet scalar field, the allowed region 
in the parameter space is divided into two classes, one of which admits Higgs bosons lighter
than $114\mbox{GeV}$ but with small couplings with the gauge bosons. We shall refer to this situation
as the light-Higgs scenario. The other class of parameters yield heavy Higgs bosons satisfying 
the present bound on the neutral scalar. This situation is called the heavy-Higgs scenario, for which
the Higgs spectrum and the phase transition are similar to those in the MSSM.\par
\begin{figure}
\centerline{\epsfig{file=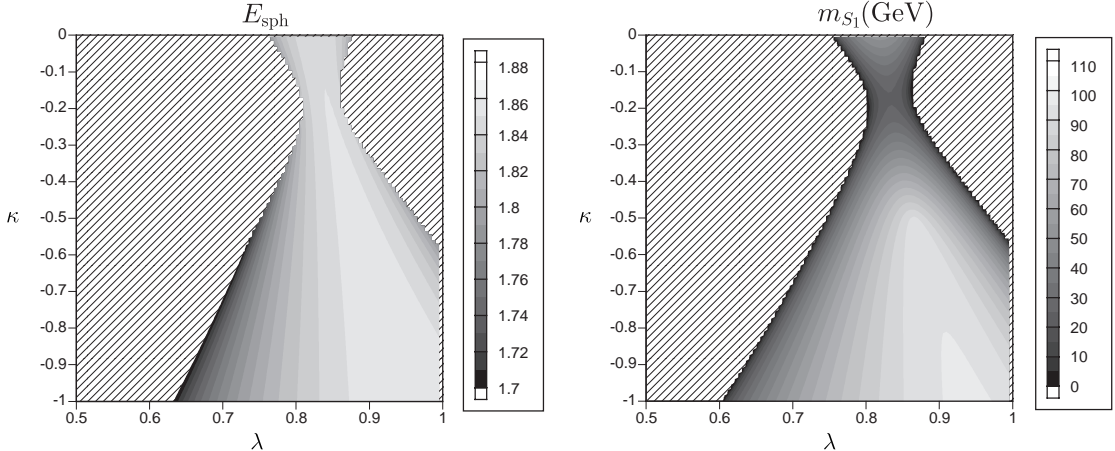, height=60mm}}
\caption{The sphaleron energy in the unit of $4\pi v_0/g_2$ and the mass of the lightest CP-even
Higgs boson as functions of $\lambda$ and $\k$ for $\tan\b=5$, $v_n=200\mbox{GeV}$, $m_{H^\pm}=600\mbox{GeV}$
and $A_\k=-100\mbox{GeV}$.}
\label{fig:mass-E}
\end{figure}

We found sphaleron solutions for wide range of the allowed parameters, as long as the smallest eigenvalue
of the mass-squared matrix of the CP-even Higgs bosons is positive, that is, the vacuum is at least a local
minimum of the potential.
This is obvious from the asymptotic behaviors of solutions, derived in the appendix.
In the heavy-Higgs scenario, the energy of the sphalerons $E$ satisfy
$1.6\le E/(4\pi v_0/g_2) \le 1.9$ for wide range of $\tan\b$ and $v_n$.
The range of the sphaleron energy coincides with that in the MSSM.
Further the energy becomes larger for larger mass of the lightest Higgs scalar. 
This behavior of the energy is expected from the results of the minimal standard model and 
those of the MSSM.
As an illustration, we show the contour plots of the sphaleron energy and the mass of the lightest
CP-even Higgs boson $m_{S_1}$ for $\tan\b=5$, $v_n=200\mbox{GeV}$, $m_{H^\pm}=600\mbox{GeV}$
and $A_\k=-100\mbox{GeV}$ in Fig.~\ref{fig:mass-E}.
For this parameter set, both the heavy-Higgs and light-Higgs scenarios are realized for different regions
in $(\lambda,\k)$-plane.
Since we are working at the tree level, some of the parameter sets may be excluded by the Higgs mass bound.
Our aim, however, is a qualitative study of the sphaleron solutions for a broad range of parameters.
The shaded regions in Fig.~\ref{fig:mass-E}, in which the smallest eigenvalue of the mass-squared matrix of 
the CP-even Higgs bosons is negative, indicates the parameters for which there is no sphaleron solution.
We found that all the solutions satisfy the condition $2c_2+c_1>0$, which is necessary for them to be locally maxima
along the noncontractible loop in the configuration space.
These contour plots show that the sphaleron energy is larger for the larger mass of the lightest Higgs scalar
in the heavy-Higgs regime ($\kappa\ltsim -0.5$).\par
\begin{figure}
\centerline{\epsfig{file=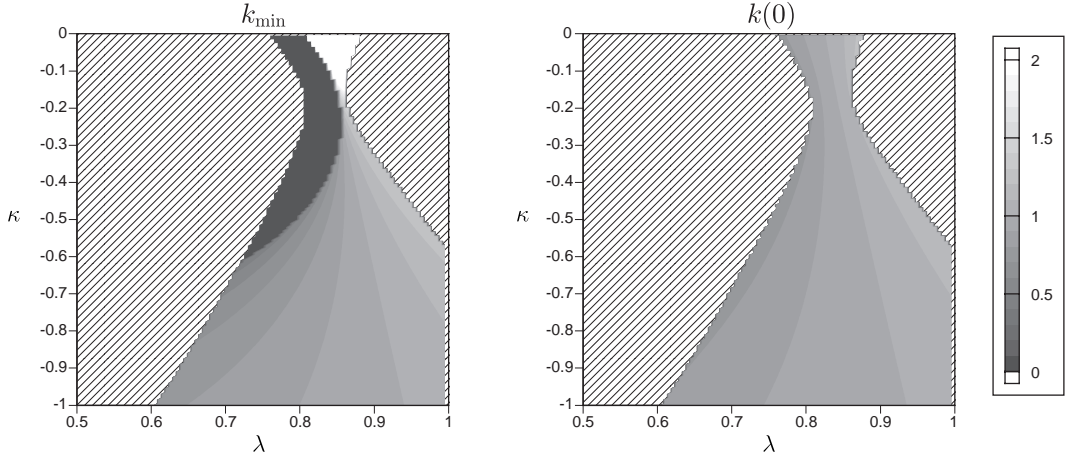, height=60mm}}
\caption{Contour plots of the location of the minimum of $U_0(k)$, $k_{\rm min}$, and the value of $k(0)$ of numerical 
solutions for the same parameter as those in Fig.~\ref{fig:mass-E}.}
\label{fig:k0}
\end{figure}
\begin{figure}
\centerline{\epsfig{file=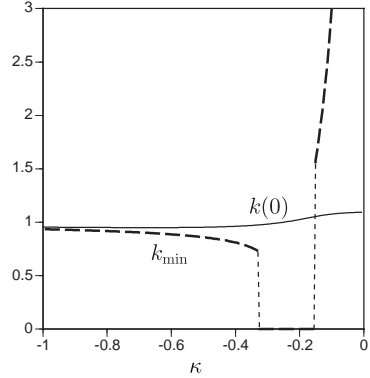, height=50mm}}
\caption{The values of $k_{\rm min}$ and $k(0)$ along the line of $\lam=0.85$ in Fig.~\ref{fig:k0}.}
\label{fig:k0-kappa}
\end{figure}
\begin{figure}
\centerline{\epsfig{file=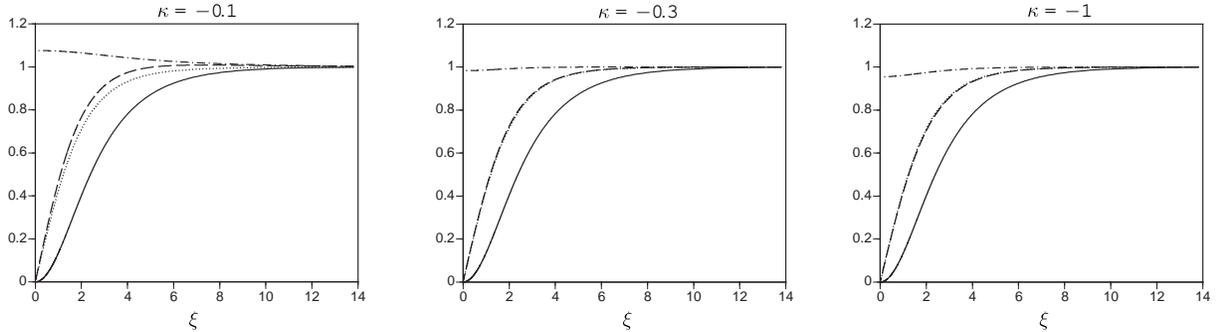, width=\textwidth}}
\caption{The profile functions $f(\xi)$ (solid curve), $h_1(\xi)$ (dashed curve), $h_2(\xi)$ (dotted curve)
and $k(\xi)$ (dashed-dottted curve) for $\k=-0.1$, $-0.5$ and $-1$ along $\lam=0.85$ 
in Fig.~\ref{fig:k0}.}
\label{fig:profiles}
\end{figure}

In the light-Higgs scenario ($-0.4\ltsim \kappa\le 0$), which is realized for small $\absv{\k}$, there is a minimum of the
potential along the $v_n$-axis. This minimum corresponds to the intermediate phase at
finite temperature, at which the electroweak gauge symmetry is restored.
Now we denote the potential in the symmetric phase ($v=0$) as $U_0(k)$ defined by
$U_0(k) = V_0(0,0, v_n k/\sqrt2)/N^2$, that is,
\begin{equation}
 U_0(k) = {{C_\k N^2}\over4}k^4 - {B\over3}k^3
 +{1\over8}\left({A\over{4N^2}}-C_\RR -2C_\lam -4C_\k N^2 + 4B\right) k^2.   \label{eq:def-U0}
\end{equation}
Depending on the parameters in the potential, $U_0(k)$ has a nontrivial minimum at $k\not=0$.
For smaller $\absv\k$, the value of $k$ at the minimum becomes larger and the value of the potential
at the minimum becomes smaller.
Although the parameters which yield deeper potential at the minimum than the prescribed vacuum are 
excluded by the vacuum condition, some of the allowed parameters admit the new type of two-stage
phase transition, which is referred to as type~B phase transition in \cite{nmssm-pt}.
As seen from Fig.~\ref{fig:mass-E}, the sphaleron energy is not so small in spite of the small mass of the lightest
Higgs scalar, in contrast to the heavy-Higgs scenario.
This small enhancement of the sphaleron energy is caused by the discrepancy of the boundary value $k(0)$ and
the location of the minimum of the potential along the $v_n$-axis.
In Fig.~\ref{fig:k0}, we show the values of $k_{\rm min}$ at which $U_0(k)$ is minimum and $k(0)$ of numerical solutions
in the $(\lam,\k)$-plane 
The difference between $k(0)$ and $k_{\rm min}$ is allowed because the boundary condition on $k(\xi)$ at $\xi=0$ is 
of Neumann type, but not of Dirichlet type.
The cross section of the contour plot at $\lam=0.85$ is depicted in Fig.~\ref{fig:k0-kappa}.
The value of $k_{\rm min}$ becomes larger than $100$ for $\k\simeq0$, while $k(0)$ stays at a value near unity.
The profile functions for $\k=-0.1$, $-0.5$ and $-1$ along $\lam=0.85$ are shown in Fig.~\ref{fig:profiles}.
The values of the sphaleron energy of these profiles are within $(1.84-1.86)\times 4\pi v_0/g_2$.
For $\k=-0.1$, $k_{\rm min}\simeq 2.99$ so that the profile with $k(0)$ near this point lowers the value
of the potential energy, while increases the gradient energy. Then the optimal profile which minimizes 
the total energy has $k(0)\simeq 1.08$.
In the heavy-Higgs scenario, where $k(\xi)\simeq 1$ for $0<\xi<\infty$, the profile is similar to
that in the MSSM.
\section{Discussions}\label{sec:discussions}
We have constructed the noncontractible loop in the configuration space of the NMSSM gauge-Higgs
fields, from which we derived the equations of motion for the sphalerons and found a necessary
condition for a solution to be a local maximum of the energy functional.
We showed that the boundary condition of the singlet scalar at the origin is of Neumann type,
in contrast to the other conditions of Dirichlet type.
This difference in the boundary conditions in the Higgs profiles is due to the lack of gauge interaction
of the singlet scalar field.\par
We numerically solved the equations of motion and found solutions for wide range of allowed
parameters of the model.
The solutions in the heavy-Higgs scenario are similar to those in the MSSM, in the sense that
$k(\xi)$ stays almost constant and the energy of the solution takes the same value as that in the MSSM.
The energy is larger for larger mass of the lightest Higgs scalar. This fact has been observed in 
the minimal standard model and in the MSSM.
The sphaleron energies in the light-Higgs scenario are almost the same as those in the heavy-Higgs scenario,
in spite of the small mass of the lightest Higgs scalar.
This is because the difference between $k(0)$ and $k_{\rm min}$, at which the potential along $v=0$
is the minimum, enhances the potential energy.
Although the Higgs potential of the NMSSM have negative contributions from the cubic terms,
the sphaleron energy in the NMSSM is almost the same as in the MSSM, for wide range of parameters.
The sphaleron energy is essential to determine the sphaleron decoupling condition after the
electroweak phase transition.
If we assume that the prefactor in the sphaleron transition rate in the NMSSM is the same
as in the MSSM, the sphaleron decoupling condition is determined solely by the energy of
the sphaleron\footnote{The prefactor is composed of the zero-mode contributions and
the factors coming from the Gaussian integrals of the positive modes.
The zero-mode contributions are mainly determined by the global symmetry of the solution,
which is common to the MSSM and the NMSSM. The positive-mode contributions depend
on the profile of the sphalerons. The effect of these prefactors on the transition rate is much
smaller than the sphaleron energy on which the rate depends exponentially.}.
Then the sphaleron decoupling condition $v_C/T_C$, where $v_C$ is the magnitude of 
the expectation value of the doublet Higgs field at the transition temperature $T_C$, also applies
to the NMSSM.\par
All the numerical solutions we have found satisfy the condition $2c_2+c_1>0$.
As seen from the definitions of $c_2$ and $c_1$, (\ref{eq:coeff-sin-mu-4}) and (\ref{eq:coeff-sin-mu-2}),
a large negative contribution is expected to come from $c_1$ when $k(\xi)$ is far from unity.
We could not find such solutions, but we cannot completely exclude the case of negative
$2c_2+c_1$. Even if we encounter the case of negative $2c_2+c_1$, it may imply
that our ansatz is inadequate for such parameters.
In that case, we will need to know the global structure of the energy functional to decide whether
the sphaleron process is possible and whether it is suppressed after the electroweak phase transition.\par
The sphaleron configurations obtained here are solutions with the tree-level potential.
As shown in \cite{mssm}, the solutions with the finite-temperature effective potential differs in the energy 
from those with the tree-level potential by several percents in the MSSM.
More precise decoupling condition of the sphaleron process in the broken phase will be obtained 
by solving the equations of motion with the effective potential with the radiative and finite-temperature
corrections in our model.
\section*{Acknowledgements}
The authors gratefully thank  S.~Otsuki for valuable discussions.
This work was supported in part by Grant-in-Aid for Scientific Research on Priority Areas No.~13135222 
to K.~F. and F.~T., and No.~17043009 to K.~F. from the MEXT of Japan.
\appendix
\section{Asymptotic solutions}\label{sec:asymptot}
We shall derive approximate solutions to the equations of motion for the sphaleron (\ref{eq:EOM-f-xi}) --
(\ref{eq:EOM-k-xi}) in the asymptotic regions at $\xi\sim0$ and $\xi\sim\infty$, respectively.
For the energy functional (\ref{eq:E-sph-xi}) to have a finite value, the profile functions must behave as
$f(\xi)\sim \xi^{1/2+\epsilon_f}$, $h_i(\xi)\sim \mbox{const.}+\xi^{-1/2+\epsilon_{h_i}}$ and
$k(\xi)\sim \mbox{const.}+\xi^{-1/2+\epsilon_k}$ at $\xi\sim0$.
Here $\epsilon$'s are some positive constants.
Similarly, all the profile functions must approach $1$ as $\xi\rightarrow\infty$.
More detailed behavior of them can be obtained by the approximate equations of motion in each
asymptotic region.
At $\xi\sim 0$, (\ref{eq:EOM-f-xi}), (\ref{eq:EOM-h1-xi}) and (\ref{eq:EOM-h2-xi}) are reduced to
\begin{eqnarray}
 && f''(\xi)\simeq 
 {2\over{\xi^2}}f(\xi)- {1\over4}\left( h_1^2(\xi)\cos^2\b+h_2^2(\xi)\sin^2\b\right),    \label{eq:asym0-f}\\
 && h''_1(\xi)+{2\over\xi}h_1(\xi)\simeq {2\over{\xi^2}}h_1(\xi), \qquad
      h''_2(\xi)+{2\over\xi}h_2(\xi)\simeq {2\over{\xi^2}}h_2(\xi),                     \label{eq:asym0-h} 
\end{eqnarray}
where we have used $f(0)=0$. 
The solutions to the second and third equations are $h_i(\xi) \propto \xi$ and $h_i(\xi) \propto \xi^{-2}$ for $i=1$ and $2$,
while the latter is excluded by the finiteness of the energy functional.
Then (\ref{eq:asym0-f}) implies that $f(\xi)\propto \xi^2$, if $f(\xi)$ yield a finite energy.
The asymptotic behavior of $k(\xi)$ is different from those of $h_1(\xi)$ and $h_2(\xi)$.
This is because the gauge interaction generates the dominant term proportional to $2/\xi^2$ in 
the right-hand sides of (\ref{eq:asym0-h}), whose counterpart does not exist in (\ref{eq:EOM-k-xi}).
In the asymptotic region $\xi\sim0$, the equation (\ref{eq:EOM-k-xi}) is approximated by
\begin{equation}
 k''(\xi) + {2\over\xi}k'(\xi) = A_0 k(\xi) -Bk^2(\xi) + C_\k N^2 k^3(\xi),    \label{eq:asym0-k}
\end{equation}
where we have put $A_0=(A\sin^2\b\cos^2\b/N^2-2C_\RR \sin\b\cos\b -2C_\lam -4C_\k N^2+4B)/4$.
In order to solve (\ref{eq:asym0-k}), we set
\begin{equation}
 k(\xi) = \gamma + \sum_{n=0}^\infty a_n \xi^{n+\nu}, \qquad (a_0\not=0, \nu>0) \label{eq:asymp-ansatz-k}
\end{equation}
where $\gamma$ is some finite constant.
Inserting this expression into (\ref{eq:asym0-k}), we obtain, to the lowest order in $\xi$,
\begin{equation}
 \nu(\nu+1)a_0\xi^{\nu-2} + (\nu+1)(\nu+2)a_1\xi^{\nu-1}+\cdots 
 = (A_0-B\gamma+C_\k N^2\gamma^2)\gamma + O(\xi^\nu).
\end{equation}
If $(A_0-B\gamma+C_\k N^2\gamma^2)\gamma=0$, this equation holds only when $\nu>2$.
When $(A_0-B\gamma+C_\k N^2\gamma^2)\gamma\not=0$, this is satisfied  if $\nu=2$ and
\begin{equation}
 6a_0 = (A_0-B\gamma+C_\k N^2\gamma^2)\gamma, \qquad a_1 = 0.
\end{equation}
Although $\gamma$ and $a_0$ are not determined unambiguously at this order, it always holds that
$k'(\xi=0)=0$.  
Therefore, we must impose the boundary conditions at $\xi=0$ of Dirichlet type for $f(\xi)$, $h_1(\xi)$ and
$h_2(\xi)$, and that of Neumann type for $k(\xi)$.\par
At $\xi\sim\infty$, all the fields must approach unity for the energy integral to be finite.
To study the asymptotic behavior, we put $f(\xi)=1+\Delta f(\xi)$, $h_i(x)=1+\Delta h_i(x)$
and $k(\xi)=1+\Delta k(\xi)$ in the equations of motion (\ref{eq:EOM-f-xi}) --
(\ref{eq:EOM-k-xi}) and keep only the linear terms in the deviations from $1$:
\begin{eqnarray}
 \Delta f''(\xi) \!\!\!&\simeq&\!\!\! \left({1\over4}+{2\over{\xi^2}}\right)\Delta f\simeq {1\over4}\Delta f(\xi),  \\
 \Delta h''_1(\xi)+{2\over{\xi^2}}\Delta h'_1(\xi)
 \!\!\!&\simeq&\!\!\!
 {A\over4}\sin^2\b(\Delta h_1-\Delta h_2 - \Delta k) + {{C_\RR N^2}\over2}\tan\b \Delta k  \nonumber\\
 &&
 +{1\over4}\left(\Delta h_1\cos^2\b - \Delta h_2\sin^2\b\right)
 +C_\lam\left(\Delta h_2\sin^2\b + N^2\Delta k\right)                    \nonumber\\
 \!\!\!&=&\!\!\!
 {{A\sin^2\b+\cos^2\b}\over4}\Delta h_1
 + \left(-{{A+1}\over4}+ C_\lam\right)\sin^2\b \Delta h_2           \nonumber\\
 &&
 + \left( -{A\over{4N^2}}\sin^2\b + {{C_\RR}\over2}\tan\b + C_\lam\right) N^2\Delta k,     \label{eq:asym-NMSSM-h1}\\
 \Delta h''_2(\xi)+{2\over{\xi^2}}\Delta h'_2(\xi) 
 \!\!\!&\simeq&\!\!\!
 {A\over4}\cos^2\b(\Delta h_2-\Delta h_1 - \Delta k) + {{C_\RR N^2}\over2}\cot\b\Delta k  \nonumber\\
 &&
 -{1\over4}\left(\Delta h_1\cos^2\b - \Delta h_2\sin^2\b\right)
 +C_\lam\left(\Delta h_1\cos^2\b + N^2\Delta k\right)                  \nonumber\\
 \!\!\!&=&\!\!\!
 \left(-{{A+1}\over4}+ C_\lam\right)\cos^2\b \Delta h_1  
 + {{A\cos^2\b+\sin^2\b}\over4}\Delta h_2               \nonumber\\
 &&
 + \left( -{A\over{4N^2}}\cos^2\b + {{C_\RR}\over2}\cot\b + C_\lam\right) N^2\Delta k, \label{eq:asym-NMSSM-h2}\\
 \Delta k''(\xi)+{2\over{\xi^2}}\Delta k'(\xi)
 \!\!\!&\simeq&\!\!\!
 {A\over{4N^2}}\sin^2\b\cos^2\b(\Delta k - \Delta h_1 - \Delta h_2)
 + {{C_\RR}\over2}\sin\b\cos\b(\Delta h_1+\Delta h_2+\Delta k )   \nonumber\\
 &&
 +C_\lam(\Delta h_1\cos^2\b + \Delta h_2\sin^2\b)
 +2C_\k N^2\Delta k -B\Delta k                \nonumber\\
 \!\!\!&=&\!\!\!
 \left( -{A\over{4N^2}}\sin^2\b + {{C_\RR}\over2}\tan\b + C_\lam\right)\cos^2\b \Delta h_1 \nonumber\\
 &&
 + \left( -{A\over{4N^2}}\cos^2\b + {{C_\RR}\over2}\cot\b + C_\lam\right)\sin^2\b \Delta h_2   \nonumber\\
 &&
 +\left( {A\over{4N^2}}\sin^2\b\cos^2\b+{{C_\RR}\over2}\sin\b\cos\b + 2C_\k N^2 -B\right)\Delta k.\label{eq:asym-NMSSM-k}
\end{eqnarray}
Now we introduce a $3\times3$ symmetric matrix $P$ whose elements are
\begin{eqnarray}
 p_{11} \!\!\!&=&\!\!\! {1\over4}(A\sin^2\b+\cos^2\b), \qquad 
 p_{12} =  \left(-{{A+1}\over4}+ C_\lam\right)\sin\b\cos\b, \nonumber\\
 p_{13} \!\!\!&=&\!\!\! \left( -{A\over{4N^2}}\sin^2\b + {{C_\RR}\over2}\tan\b + C_\lam\right)N\cos\b ,\qquad
 p_{22} = {1\over4}(A\cos^2\b+\sin^2\b),      \nonumber\\
 p_{23} \!\!\!&=&\!\!\! \left( -{A\over{4N^2}}\cos^2\b + {{C_\RR}\over2}\cot\b + C_\lam\right)N\sin\b,   \nonumber\\
 p_{33} \!\!\!&=&\!\!\! {A\over{4N^2}}\sin^2\b\cos^2\b+{{C_\RR}\over2}\sin\b\cos\b + 2C_\k N^2 -B.
\end{eqnarray}
The matrix $P$ is related to the tree-level mass-squared matrix of the CP-even Higgs bosons ${\mathcal M}_S^2$
by $P = {\mathcal M}_S^2/(4m_W^2)$.
As long as we restrict the parameter sets to satisfy the vacuum condition, all the eigenvalues of $P$ are
positive definite.
Then Eqs.~(\ref{eq:asym-NMSSM-h1}) -- (\ref{eq:asym-NMSSM-k}) are written as
\begin{equation}
 {{d^2}\over{d\xi^2}}
 \pmatrix{ \Delta\tilde h_1(\xi)\cos\b \cr \Delta\tilde h_2(\xi)\sin\b \cr N\Delta\tilde k(\xi) } \simeq
 P\pmatrix{ \Delta\tilde h_1(\xi)\cos\b \cr \Delta\tilde h_2(\xi)\sin\b \cr N\Delta\tilde k(\xi) },
\end{equation}
where we have defined $\Delta\tilde h_i(\xi)=\xi\Delta h_i(\xi)$ and $\Delta\tilde k(\xi)=\xi\Delta k(\xi)$.
If we denote the eigenvalue of $P$ as $\lambda_a$, $\Delta\tilde h_i(\xi)$ and $\Delta\tilde k(\xi)$
are expressed as linear combinations of $e^{-\sqrt{\lambda_a}\xi}=e^{-m_{S_a}r}$, where
$m_{S_a}$ is the tree-level mass of the CP-even Higgs boson.
Hence the asymptotic behaviors of the solutions are governed by the smallest mass of the CP-even Higgs boson.


%
%

\end{document}